\documentclass[english]{article}
\usepackage[T1]{fontenc}
\usepackage[utf8]{inputenc}
\usepackage{geometry}
\usepackage{babel}
\usepackage{amsmath}
\usepackage{natbib}

\title{Thermodynamically consistent large-eddy simulation models}

\author{Thomas Dubos \\ Laboratoire de Météorologie Dynamique/IPSL, École Polytechnique \\ 91120 Palaiseau, France}

\begin{document}
\maketitle

\begin{abstract}

Filtered budgets for anelastic turbulence and a general expression
of the turbulent sensible heat flux are derived for a multicomponent
fluid with an arbitrary equation of state. A family of subgrid-scale
closures is then found under the constraint of consistency with (i)
the first and second laws of thermodynamics and (ii) invariance with
respect to irrelevant thermodynamic constants. A similar family of fully compressible models is also constructed heuristically. These models predict
turbulent kinetic energy, assume down-gradient closures for three-dimensional
turbulent fluxes and impose certain relationships between the closures
for the turbulent fluxes of heat, matter, entropy, and the work
of buoyancy forces.

A key finding is the explicit derivation of the local rate of entropy
production in the filtered model. Positive entropy production is
guaranteed whenever the turbulent diffusions of heat and composition
are positive and no cross-diffusion occurs. Cross-diffusivities are
admissible provided their magnitude is within the bounds of an explicit
criterion. The filtered model is invariant under a wider class of
transformations than the unfiltered model. Furthermore, in the special
case of a single turbulent diffusivity, an arbitrary conservative
variable can be prognosed while ignoring its precise relationship
to entropy.

These findings show that down-gradient closures are consistent
with the first and second law of thermodynamics even when they lead
to a turbulent sensible heat flux up the temperature gradient. Indeed, while molecular conduction/diffusion is spontaneous and energy-conserving, stratified turbulent mixing is driven by mechanical turbulence and enabled by the consumption of
turbulent kinetic energy.

\end{abstract}


\section{Introduction}

The study of many turbulent flows
relies on models whose aim is to explicitly describe the flow
only down to a certain scale, omitting finer-scale details or, more
precisely, taking them into account only in a statistical sense. Such
models are, at least conceptually, obtained through a coarse-graining
procedure, which can take the form of averaging or filtering. Among
such coarse-grain models, large-eddy simulation (LES) models aim at
explicitly representing scales down to a cut-off scale assumed to
lie within the inertial range of the direct turbulent energy cascade,
thus ensuring that most of the kinetic energy of the flow is resolved
\citep{sagaut_large_2006}. For atmospheric or oceanic flows, this
cut-off scale is typically on the order of meters or tens of meters
\citep{lilly_numerical_1962,deardorff_three-dimensional_1974,moeng_evaluation_1989,pressel_large-eddy_2015,chammas_accelerating_2023}. 
Coarse-graining the Navier-Stokes equations \citep{groot_non-equilibrium_1962}
or other similar equations such as the anelastic or Boussinesq systems
\citep{eldred_thermodynamically_2021,tailleux_simple_2024} leads to
the appearance of turbulent fluxes in the coarse-grained model. A
key task of turbulence modelling is to assign a value to these fluxes
via turbulent closures, which take as inputs the resolved flow quantities
and possibly additional sub-grid statistics also predicted by the
model, such as turbulent kinetic energy. For LES, a widely used approach
is to model turbulent fluxes of momentum and transported scalars as
``down-gradient'', i.e. proportional and opposed to the gradient
of momentum and concentration \citep{lilly_numerical_1962,deardorff_three-dimensional_1974,moeng_evaluation_1989,pressel_large-eddy_2015,chammas_accelerating_2023}.
The mixing coefficient itself remains to be specified, and many approaches
have been proposed \citep{sagaut_large_2006}.\\

When the flow of interest is at a low Mach number and the effect of
gravity is negligible, the relevant unfiltered model is the incompressible
Navier-Stokes model \citep{sagaut_large_2006}. In this model, thermodynamics
decouple from dynamics, so that temperature becomes a passive scalar.
The relevant energy budget is that of kinetic energy, which is not
closed: kinetic energy, in the absence of forcings, decreases due
to viscous dissipation. In a real fluid, dissipation of kinetic energy
is compensated by an increase of internal energy and thus of temperature,
but in the incompressible Navier-Stokes model this temperature increase
does not feedback onto the flow and is dynamically irrelevant. From
the point of view of an LES model, dissipation occurs through viscosity
and diffusion at an unspecified, unresolved scale and turbulence provides
a route to dissipation which starts by the conversion between resolved
kinetic energy and turbulent (unresolved) kinetic energy (TKE). Since
TKE is ultimately dissipated after cascading to small scales, closures
are expected to decrease resolved kinetic energy on average. This
expectation is more or less the only energetic constraint on closures,
and is met by down-gradient closures.\\

Geophysical flows are, by definition, subject to gravity (and possibly
rotation), and are typically stratified. As a consequence, their dynamics
is coupled to their thermodynamics unlike in the incompressible Navier-Stokes
model, and their energetics are richer in several ways. The flow description
itself must include some form of temperature or entropy, such as potential
temperature. This scalar is transported but not passively, since it
controls the buoyancy force felt by fluid parcels. Furthermore its
turbulent fluxes are related to turbulent heat transfer. Non-kinetic energy
now includes geopotential energy \citep{tort_usual_2014}, and conversions
between kinetic and potential energy (understood here as internal+geopotential
energy) are significant. Whereas mixing in a homogeneous fluid is
energy-neutral, mixing a stably stratified environment increases potential
energy and consumes kinetic energy \citep{peltier_mixing_2003}. Furthermore,
due to the strong pressure gradient caused by gravity, temperature
is distinct from potential temperature and other entropy-like quantities.
Since it is the latter that are typically assumed to ``be mixed''
by turbulent closures, this leads to the possibility of turbulent
heat fluxes transferring energy from cold fluid to warmer fluid, apparently
contradicting the second law of thermodynamics \citep{gassmann_entropy_2018,gassmann_entropy_2019}.\\

In geophysical flows, the fluid density may vary substantially
in space due to gravity. Except perhaps on certain exoplanets, geophysical flows are also low-Mach-number
flows. As a result, they are nearly-incompressible in the sense that their density at a given position does not significantly respond to pressure fluctuations resulting from the fluid motion itself. The anelastic model \citep{ogura_scale_1962,lipps_scale_1982,bannon_anelastic_1996,pauluis_thermodynamic_2008}
is a relevant fine-grain model in this context, and bears many similarities
to the incompressible Navier-Stokes model. However, until recently,
there was no general self-consistent anelastic model from which turbulent
budgets could be derived exactly through a coarse-graining procedure.
More precisely, a model of anelastic flow including viscosity and
diffusion, valid for arbitrary thermodynamics, including multicomponent
fluids (e.g. moist air and seawater) and an arbirary density profile,
consistent with the first and second laws of thermodynamics, and with
energetics transparently related to those of the compressible Navier-Stokes
model, was not available until recently \citep{eldred_thermodynamically_2021,tailleux_simple_2024}. 

Geophysical LES models often complement the resolved flow equations
with a predictive equation for turbulent kinetic energy , or a more
detailed description of unresolved velocity correlations. One motivation
is to have TKE enter as an input when determining turbulent mixing
coefficients. Another motivation is to provide a more detailed description
of the route of energy to dissipation, which can provide energetic
constraints on closures. Since they were derived before the recent
generalizations of anelastic models, turbulent budgets used by current
LES models are either rigorously derived but incorporate restrictions
that can now be relaxed \citep{stull_introduction_1988,pressel_large-eddy_2015}
or careful but heuristic extensions of the latter. \\

Turbulent closures should be as \emph{accurate} as possible across a broad range of conditions. Additionally, it is crucial that they are \emph{consistent}, a concept that involves multiple dimensions.
For instance, certain closures used in incompressible turbulence must obey so-called realizability conditions \citep{schumann_realizability_1977, sagaut_large_2006}. For closures used in geophysical models, it remains unclear to which extent their
form is restricted by energetics (first law of thermodynamics) and
by the requirement that unresolved dissipation increase entropy (second
law of thermodynamics) \citep{lauritzen_reconciling_2022}. Recently, the author has raised the so far overlooked issue that the observable predictions made by a coarse-grained model should be
independent from those thermodynamic constants that are irrelevant
for the fine-grain model, in the sense that changing them does not
alter its observable predictions \citep{dubos_thermodynamic_2024}.
This work has led to some insights into the construction of inputs
and outputs of closures. Following this first step, the main goal
of this work is (i) to set on firm ground the various turbulent budgets
that LES models use or should respect and (ii) examine the consistency
of down-gradient closures with these budgets and the first and second
laws of thermodynamics. \\

To this end, section 2 provides the necessary background on thermodynamics
and anelastic dynamics of a multicomponent fluid. Section 3 uses a
coarse-graining procedure to derive various exact turbulent budgets.
Especially, a general, explicit and thermodynamically-invariant definition
of the turbulent sensible heat flux is obtained. Section 4 introduces
an incompletely specified anelastic LES model. Relationships between
certain closures are derived in order to satisfy the first law. Closing
the energy budget is a prerequisite to obtain an expression for the
local production of entropy, a key result of this work. This expression
is closely related to the expression for entropy produced during isobaric
mixing of fluid parcels. Section 5 discusses the invariance properties
of the LES model, especially in the case where a single turbulent
diffusivity is used for temperature and composition. In section 6,
the anelastic assumption is removed heuristically, leading to thermodynamically-consistent
compressible LES models. Section 7 discusses the findings and concludes.

\section{Background}

\subsection{Thermodynamics}

We consider a binary fluid \citep{ooyama_thermodynamic_1990,bannon_hamiltonian_2003}.
A fluid parcel contains masses $m_{d}$ and $m_{w}$ of the each constituent,
e.g. dry air and water for moist air, or freshwater and salt for seawater.
Here we consider only two constituents, but it is straightforward
to consider an arbitrary number of constituents. Thermodynamics derive
from an unspecified specific Gibbs function $g(p,T,q)$ with $p,T$
pressure and temperature and $q=m_{w}/(m_{w}+m_{d})$ specific humidity
(or absolute salinity for seawater). Derivatives $\partial f(p,T,q)/\partial X$,
$X=p,T,q$ are noted\footnote{Superscripts are used here because subscripts will be used later for other partial derivatives} $f^{X}$ . From $g(p,T,q$) one recovers specific
volume $\upsilon=g^{p}$, specific entropy $s=-g^{T}$, specific enthalpy
$h=g+Ts$, as well as partial enthalpies $h^{w}=h+(1-q)h^{q},\,h^{d}=h-qh^{q}$
(and similarly for entropy) so that:
\begin{align}
h^{q} & =h^{w}-h^{d}, & h & =qh^{w}+(1-q)h^{d}\\
s^{q} & =s^{w}-s^{d}, & s & =qs^{w}+(1-q)s^{d}\\
g^{q} & =\mu^{w}-\mu^{d}=h^{q}-Ts^{q} & g & =q\mu^{w}+(1-q)\mu^{d}
\end{align}
with $s^{w},\,s^{d}$ partial entropies and $\mu^{w},\,\mu^{d}$ chemical
potentials \citep{bannon_hamiltonian_2003}.

\subsection{Thermodynamic invariance}

For a model quantity $X$, \cite{dubos_thermodynamic_2024} introduces
the notion of the infinitesimal variation $\delta X$ at constant
$p,T,q$, which are the observable state variables. Such a change
can be especially due to a change of reference enthalpies and entropies,
resulting from defining $g(p,T,q)$ from another Gibbs function $g^{*}(p,T,q)$
as:

\begin{align*}
g(p,T,q) & =g_{*}(p,T,q)+(1-q)\left(\delta h_{0}^{d}-T\delta s_{0}^{d}\right)+q\left(\delta h_{0}^{w}-T\delta s_{0}^{w}\right)
\end{align*}
where $\delta h_{0}^{d}$, $\delta h_{0}^{w}$ $\delta s_{0}^{d},\,\delta s_{0}^{w}$
are arbitray constants. The entropy and enthalpy defined by $g$ do
depend on reference values :
\begin{align}
\delta s & =(1-q)\delta s_{0}^{d}+q\delta s_{0}^{w}\\
\delta h & =(1-q)\delta h_{0}^{d}+q\delta h_{0}^{w}
\end{align}
where $\delta X$ compares $X$ obtained from $g$ to $X$ obtained
from $g^{*}$. 

Model constants are declared \emph{irrelevant} if $\delta X=0$ for
any directly observable model quantity (such as $p,T,q)$ and their
model-predicted time derivative. In the Navier-Stokes, pseudo-incompressible
and anelastic models with a consistent treatment of molecular diabatic
processes \citep{groot_non-equilibrium_1962,pauluis_thermodynamic_2008,eldred_thermodynamically_2021,tailleux_simple_2024},
reference enthalpies and entropies are irrelevant, except in the presence
of chemical reactions and phase change. It is therefore desirable
to identify and manipulate quantities $X$ that are \emph{invariant},
in the sense that $\delta X=0$. 

Now consider an infinitesimal displacement in $(p,T,q)$ space. Changes
in entropy and enthalpy $s^{q}\text{d}q$ and $h^{q}\text{d}q$ are
due purely to a change in composition. Thus the \emph{reduced} increments:
\begin{align*}
\tilde{\text{d}}s & \equiv\text{d}s-s^{q}\text{d}q=q\text{d}s^{w}+(1-q)\text{d}s^{d}\\
\tilde{\text{d}}h & \equiv\text{d}h-h^{q}\text{d}q=q\text{d}h^{w}+(1-q)\text{d}h^{d}
\end{align*}
are variations at constant composition, and are invariant. Indeed
$\delta s^{w}=\delta s_{0}^{w}=cst$ so that $\delta\text{d}s^{w}=0$.
Since $\delta q=0$, $\delta\tilde{\text{d}}s=0$. In the sequel we
will use similarly defined reduced gradients and reduced turbulent
fluxes \citep{dubos_thermodynamic_2024}. 

\subsection{Anelastic motion}

For simplicity, Cartesian coordinates $(x,y,z)$ are used, $z$ being
vertical and upwards, i.e. geopotential is $\Phi=g_{grav}z$ with
gravity $g_{grav}$ a constant. A function $f(p,s,q)$ can also be
regarded as a field $f(t,x,y,z)$ where $t$ is time. In this case
its derivatives are noted $\nabla_{i}f$, $i=t,x,y,z$. Especially
the material (Lagrangian) derivative is $D_{t}=\nabla_{t}+\mathbf{u}\cdot\nabla$
with $\mathbf{u}=(u,v,w)$ the fluid velocity.

In the equations of motion, it is more convenient to regard enthalpy
as a function of its canonical variables $(p,s,q)$ rather than $(p,T,q)$.
To distinguish the derivatives of a thermodynamic function $f(p,s,q)$
from those of $f(p,T,q)$, we note $f_{X}=\partial f(p,s,q)/\partial X$
$(X=p,s,q)$. For instance, from $\text{d}h=\upsilon\,\text{d}p+T\text{d}s+h_{q}\text{d}q$
on finds:
\begin{equation}
h^{q}=h_{q}+Ts^{q}\label{eq:hq}
\end{equation}

In anelastic systems, thermodynamic functions are evaluated at $(p_{ref}(z),s,q)$
with $p_{ref}(z)$ a reference pressure profile, to which hydrostatic
balance associates a density profile $\rho_{ref}$ by $\text{d}p_{ref}=-g_{grav}\,\rho_{ref}\text{d}z$.
Specific enthalpy $h(p,s,q)$ and its derivatives become functions
of $z,s,q$. For such functions we also note $f_{X}=\partial f(z,s,q)/\partial X$
$(X=z,s,q)$. 

As stated in the introduction, restrictions on anelastic systems such
as the need to linearize the equation of state or to consider only
an adiabatic reference profile \citep{pauluis_thermodynamic_2008}
have been fully relaxed \citep{eldred_thermodynamically_2021}. The
equivalent formulation by \cite{tailleux_simple_2024} is followed
here. In this formulation, a central role is played by static energy
$\Sigma$:
\begin{equation}
\Sigma(z,s,q)=\Phi(z)+h(z,s,q).
\end{equation}
 In \cite{tailleux_simple_2024}, anelastic static energy is defined
with an extra term, linear in the pressure in excess of $p_{ref}(z)$,
which acts as a Lagrange multiplier. Since this extra term vanishes
when the anelastic constraint $\rho=\rho_{ref}$ is satisfied, it
is omitted here. Buoyancy $b$ and Brunt-Vaisala frequency derive
from $\Sigma$ as:
\begin{equation}
b(z,s,q)=g_{grav}\left(\upsilon\rho_{ref}-1\right)=-\Sigma_{z}
\end{equation}
\begin{align}
N^{2} & =\nabla_{z}b-b_{z}=-h_{zs}\nabla_{z}s-h_{zq}\nabla_{z}q\label{eq:BruntVaisala}
\end{align}
where $\upsilon$ is specific volume evaluated at $p_{ref}(z),s,q$. 

The anelastic equations of motion are then :
\begin{align}
\rho-\rho_{ref} & =0\label{eq:AN_EOS}\\
\nabla_{t}\rho+\nabla\cdot\rho\mathbf{u} & =0\label{eq:AN_mass}\\
\nabla_{t}\left(\rho q\right)+\nabla\cdot\rho q\mathbf{u} & =-\nabla\cdot\mathbf{j}_{q}\label{eq:composition}\\
\nabla_{t}\left(\rho s\right)+\nabla\cdot\rho s\mathbf{u} & =\sigma_{s}-\nabla\cdot\mathbf{j}_{s}\label{eq:AN_entropy}\\
D_{t}\mathbf{u}+\nabla\frac{\pi}{\rho} & =b\nabla z+\frac{1}{\rho}\nabla\cdot\boldsymbol{\tau}\label{eq:AN_momentum}
\end{align}
with $\rho$ the fluid mass per unit volume, $\pi$ the pressure in
excess of $p_{ref}$, $\mathbf{j}_{s}$ and $\mathbf{j}_{q}$ the
fluxes of entropy and composition due to conduction and diffusion,
$\sigma_{s}$ the entropy production rate and $\boldsymbol{\tau}$
the viscous stress tensor.

\subsection{Anelastic energetics}

Let $\varepsilon=\nabla\mathbf{u}:\boldsymbol{\tau}$ be the viscous
dissipation rate and: 
\begin{equation}
\mathbf{j'}_{s}\equiv\mathbf{j}_{s}-s^{q}\mathbf{j}_{q},\qquad\mathbf{j}_{H}=T\mathbf{j}'_{s}\label{eq:reduced_fluxes}
\end{equation}
with $\mathbf{j'}_{s}$ the reduced entropy flux and $\mathbf{j}_{H}$
the sensible heat flux. In order to ensure conservation of energy,
the entropy source $\sigma_{s}$ is the sum of a source due to viscous
dissipation and another source due to conduction and diffusion \citep{groot_non-equilibrium_1962,eldred_thermodynamically_2021,tailleux_simple_2024}:
\begin{equation}
T\sigma_{s}=\varepsilon+T\sigma_{mix},\qquad T\sigma_{mix}=-\mathbf{j'}_{s}\cdot\nabla T+\mathbf{j}_{q}\cdot\left(\nabla\mu-\mu^{T}\nabla T\right)\label{eq:entropy_source}
\end{equation}
with temperature $T=h_{s}$ and chemical potential $\mu=h_{q}=g^{q}=h^{w}-h^{d}-T\left(s^{w}-s^{d}\right)$.

The budgets of kinetic, potential and total energy ensuing from (\ref{eq:AN_EOS}-\ref{eq:AN_momentum})
are then:
\begin{align}
\nabla_{t}\rho K+\nabla\cdot\left(\rho K\mathbf{u}+\pi\mathbf{u}-\mathbf{u}\cdot\boldsymbol{\tau}\right) & =\rho b\mathbf{u}\cdot\nabla z-\varepsilon\label{eq:AN_KE_budget}\\
\nabla_{t}\left(\rho\Sigma-p_{ref}\right)+\nabla\cdot\left(\rho\Sigma\mathbf{u}+h^{q}\mathbf{j}_{q}+\mathbf{j}_{H}\right) & =\varepsilon-\rho b\mathbf{u}\cdot\nabla z\label{eq:AN_PE_budget}\\
\nabla_{t}E+\nabla\cdot\left[\left(E+p_{ref}+\pi\right)\mathbf{u}+\mathbf{j}_{E}\right] & =0
\end{align}
with $K=\mathbf{u}\cdot\mathbf{u}/2$ specific kinetic energy, $E=\rho\left(K+\Sigma\right)-p_{ref}$
total energy per unit volume, $\left(E+p_{ref}+\pi\right)\mathbf{u}$
the adiabatic energy flux and 
\begin{equation}
\mathbf{j}_{E}=h^{q}\mathbf{j}_{q}+\mathbf{j}_{H}-\mathbf{u}\cdot\boldsymbol{\tau}\label{eq:irreversible_energy_flux}
\end{equation}
 the energy flux due to viscosity, conduction and diffusion. 

Note that specific entropy $s$, total energy $E$, and the fluxes
$\mathbf{j}_{s},\,\mathbf{j}_{E}$ are not invariant. However the
reduced entropy flux $\mathbf{j'}_{s}$ and sensible heat flux $\mathbf{j}_{H}$
are invariant, as well as $\nabla\mu-\mu^{T}\nabla T=g^{qp}\nabla p_{ref}+g^{qq}\nabla q$ and the reduced entropy gradient \citep{dubos_thermodynamic_2024}:
\begin{equation}
\tilde{\nabla}s = \nabla s - s^q \nabla q = -g^{TT} \nabla T - g^{pT}\nabla p_{ref}
\label{eq;reduced_entropy_flux}.
\end{equation}
Thus $\sigma_{s}$ is an invariant
expression, bilinear in the fluxes and the so-called thermodynamic
forces $\nabla\mathbf{u}$, $\nabla T$, $\nabla\mu-\mu^{T}\nabla T$. 

The anelastic system is closed by introducing phenomenological laws
- not explicitly needed for the present purposes - yielding $\boldsymbol{\tau}$,
$\mathbf{j}'_{s}$ and $\mathbf{j}{}_{q}$ from thermodynamic forces.
A linear elliptic problem for $\pi$ can then be formulated.

\subsection{Entropy production in the turbulent limit\label{subsec:Entropy-production-turbulent}}

It will be useful in the next section to estimate entropy production
by mixing to leading order in the turbulent limit. This limit is
defined here by the assumption that the irreversible fluxes $\mathbf{j'}_{s}$,
$\mathbf{j}_{q}$ and $\boldsymbol{\tau}$ become negligble compared
to turbulent fluxes but simultaneously the gradients $\nabla s,\,\nabla q,\nabla\mathbf{u}$
become large (due to stirring), so that $\varepsilon$ and $\sigma_{mix}$
remain finite. Since $p_{ref}(z)$, on the other hand, remains unchanged
and finite: 
\begin{align}
\tilde{\nabla}s & \rightarrow-g^{TT}\nabla T\nonumber \\
\nabla\mu-\mu^{T}\nabla T & \rightarrow g^{qq}\nabla q\nonumber \\
T\sigma_{mix} & \rightarrow T\sigma_{iso}\equiv\left(g^{TT}\right)^{-1}\mathbf{j'}_{s}\cdot\tilde{\nabla}s-g^{qq}\mathbf{j}_{q}\cdot\nabla q\label{eq:sigma_iso}
\end{align}

More insight is gained by relating this expression to enthalpy. To
this end, we consider the process of \emph{isobaric mixing, }defined
as the process leading two fluid parcels of unit mass with state $(p,s\pm\text{d}s,q\pm\text{d}q)$
to a final equilibrium state $(p,s+\Delta s_{iso},q)$. During this
irreversible isobaric process, specific entropy increases by $\Delta s_{iso}$
while enthalpy is conserved :
\[
T\Delta s_{iso}=\frac{h\left(p,s+\text{d}s,q+\text{d}q\right)+h\left(p,s-\text{d}s,q-\text{d}q\right)}{2}-h\left(p,s,q\right)=\text{d}^{2}h>0.
\]
From $T=h_{s}$ and $\mu=g^{q}=h_{q}$, we obtain $\text{d}T=h_{ss}\text{d}s+h_{qs}\text{d}q$
and $\text{d}\mu=h_{qq}\text{d}q+h_{qs}\text{d}s=g^{qq}\text{d}q+g^{qT}\text{d}T$
hence $h_{qs}=g^{qT}h_{ss}$, $h_{qq}=g^{qq}+g^{qT}h_{qs}$. Using
$\text{d}s=-g^{TT}\text{d}T-g^{qT}\text{d}q$ yields $g^{TT}=-h_{ss}^{-1}$.
Letting $\text{d}s=\tilde{\text{d}}s-g^{qT}\text{d}q$ in $\text{d}^{2}h$
yields : 

\begin{equation}
\text{d}^{2}h= \frac{1}{2}h_{ss}\text{d}s^{2}+h_{qs}\text{d}s\text{d}q+\frac{1}{2}h_{qq}\text{d}q^{2} 
= \frac{1}{2}h_{ss}\tilde{\text{d}}s^{2}+\frac{1}{2}g^{qq}\text{d}q^{2}
\end{equation}

Thus the bilinear form $T\sigma_{iso}((\mathbf{j'}_{s},\mathbf{j}_{q}),\,(\tilde{\nabla}s,\nabla q))$
derives directly from the quadratic form $\text{d}^{2}h(\tilde{\text{d}}s,\text{d}q)$
and $\sigma_{iso}$ is the entropy source due to \emph{isobaric} conduction/diffusion. 

The underlying physics is very simple: in the turbulent limit, the
pressure gradient remains finite while the gradients of entropy and
composition become arbitrarily large. Therefore, local mixing processes
become effectively isobaric.

\section{Filtered anelastic dynamics and energetics}

\subsection{Filtering}

A filtering operation is now introduced that eliminates small-scale
details, e.g. to a field $X$ one associates the filtered field $\overline{X}$.
In theoretical developments one often considers that $X\mapsto\overline{X}$
is an ensemble average that satisfies the criterion of Reynolds averaging,
especially:
\begin{equation}
\overline{XY}\,=\overline{X}\,\overline{Y}+\overline{X'Y'}\qquad\text{where}\qquad X'=X-\overline{X}\label{eq:Reynolds}
\end{equation}
In the context of large-eddy simulations, it is typical to define
this operation rather as a convolution by some smoothing kernel, which
violates the Reynolds condition $\overline{\overline{X}}=\overline{X}$
from which (\ref{eq:Reynolds}) is derived \citep{sagaut_large_2006}.
Thus (\ref{eq:Reynolds}) is replaced by: 
\begin{equation}
\overline{XY}\,=\overline{X}\,\overline{Y}+\left(\overline{XY}-\overline{X}\,\overline{Y}\right)\label{eq:LES}
\end{equation}
In the sequel, the so-called sub-filter quantity $\overline{XY}-\overline{X}\,\overline{Y}$
is to be replaced by a closure relationship. Therefore, we can without
consequence make the abuse of notation to note it as $\overline{X'Y'}$.

\subsection{Reduced turbulent fluxes}

Folllowing \cite{dubos_thermodynamic_2024}, let us examine how turbulent fluxes vary when irrelevant thermodynamic constants change. Since $\delta s=(1-q)\delta s^{d}+q\delta s^{w}$ with $\delta s^{d}$
and $\delta s^{w}$ constants:

\begin{align}
\delta\overline{s'\mathbf{u}'} & =\delta\overline{s^{q}}\,
\end{align}
A similar identity holds for $\overline{h'\mathbf{u}'}$. Hence,
 invariant reduced 
 fluxes $\widetilde{s'\mathbf{u}'}$, $\widetilde{h'\mathbf{u}'}$
are defined by:
\begin{align}
\overline{s'\mathbf{u}'} & =\overline{s^{q}}\,\overline{q'\mathbf{u}'}+\widetilde{s'\mathbf{u}'}\label{eq:invariant_flux}\\
\overline{h'\mathbf{u}'} & =\overline{h^{q}}\,\overline{q'\mathbf{u}'}+\widetilde{h'\mathbf{u}'}
\end{align}

\subsection{Filtered dynamics}

Filtering over non-turbulent quantities are now omitted , i.e. $u$
stand for $\overline{u}$. Filtering the equations of motion yields
:
\begin{align}
\rho & =\rho_{ref}\\
\nabla_{t}\rho+\nabla\cdot\rho\mathbf{u} & =0\\
\rho D_{t}q+\nabla\cdot\rho\,\overline{q'\mathbf{u}'} & =0\\
\rho D_{t}s+\nabla\cdot\rho\,\overline{s'\mathbf{u}'} & =\sigma_{s}\\
\rho D_{t}u+\nabla\cdot\rho\overline{u'\mathbf{u}'}+\nabla_{x}\frac{\pi}{\rho} & =0\\
\rho D_{t}v+\nabla\cdot\rho\overline{v'\mathbf{u}'}+\nabla_{y}\frac{\pi}{\rho} & =0\\
\rho D_{t}w+\nabla\cdot\rho\overline{w'\mathbf{u}'}+\rho\nabla_{z}\frac{\pi}{\rho} & =b
\end{align}
where irreversible (viscous, conductive, diffusive) fluxes are neglected
but not entropy production $\sigma_{s}$. Put together, $(\rho\overline{u'\mathbf{u}'},\,\rho\overline{v'\mathbf{u}'},\,\rho\overline{w'\mathbf{u}'})$
form the Reynolds stress tensor $\mathbf{R}=\rho\overline{\mathbf{u}'\otimes\mathbf{u}'}$.

\subsection{Filtered energetics}

Filtering the budgets (\ref{eq:AN_KE_budget}-\ref{eq:AN_PE_budget})
of KE and potential energy yields :
\begin{align}
\rho D_{t}K+\nabla\cdot\mathbf{J}_{K} & =\rho b\mathbf{u}\cdot\nabla z-P_{shear}\label{eq:KE_budget}\\
\rho D_{t}k+\nabla\cdot\mathbf{J}_{k} & =P_{shear}+\rho\overline{b'w'}-\varepsilon\label{eq:TKE_budget}\\
\nabla_{t}(\rho\Sigma-p_{ref})+\nabla\cdot\rho\Sigma\mathbf{u}+\nabla\cdot\rho\overline{h'\mathbf{u}'} & =-\rho b\mathbf{u}\cdot\nabla z-\rho\overline{b'w'}+\varepsilon\label{eq:enthalpy_budget}
\end{align}
where $k=\overline{\mathbf{u}'\cdot\mathbf{u}'}/2$ is TKE and:
\begin{align}
P_{shear} & \equiv\rho\left(\overline{u'\mathbf{u}'}\cdot\nabla u+\overline{v'\mathbf{u}'}\cdot\nabla v+\overline{w'\mathbf{u}'}\cdot\nabla w\right)=\mathbf{R}:\nabla\mathbf{u}\\
\mathbf{J}_{K} & \equiv\rho\left(u\,\overline{u'\mathbf{u}'}+v\,\overline{w'\mathbf{u}'}+w\,\overline{w'\mathbf{u}'}\right)=\mathbf{R}\cdot\mathbf{u}\\
\mathbf{J}_{k} & \equiv\frac{1}{2}\rho\overline{\left(u'^{2}+v'^{2}+w'^{2}\right)\mathbf{u}'}+\overline{\pi'\mathbf{u}'}
\end{align}
Notice that in (\ref{eq:enthalpy_budget}), the source of potential
energy includes, in addition to viscous dissipation, (minus) the buoyant
production of TKE $\overline{b'w'}$. Indeed turbulent mixing, unlike
molecular conduction/diffusion, is not in general potential-energy-neutral
: over a stably stratified background, turbulent mixing tends to increase
potential energy, which requires a supply of energy (taken from TKE,
itself converted from resolved KE by shear at rate $P_{shear}$) while
over a statically unstable background, it releases energy (as TKE)
and can thus occur spontaneously.

We now decompose the turbulent enthalpy flux as $\rho\overline{h'\mathbf{u}'}=\rho h^{q}\overline{q'\mathbf{u}'}+\mathbf{J}_{H}$
with
\begin{equation}
\mathbf{J}_{H}=\rho\widetilde{h'\mathbf{u}'}.\label{eq:sensible_heat_flux}
\end{equation}
Summing up (\ref{eq:KE_budget}-\ref{eq:enthalpy_budget}) one finds
that the flux of total energy $E+\rho k$ (including TKE) is:
\begin{equation}
\mathbf{J}_{E}=\rho h^{q}\overline{q'\mathbf{u}'}+\mathbf{J}_{H}+\mathbf{u}\cdot\mathbf{R}+\mathbf{J}_{k}\label{eq:total_energy_flux}
\end{equation}
Expression (\ref{eq:total_energy_flux}) has the same structure as
(\ref{eq:irreversible_energy_flux}), except for the last term $\mathbf{J}_{k}$
which is genuinely turbulent. This similarity allows us to recognize
$\mathbf{J}_{H}$ unambiguously as the \emph{turbulent sensible heat
flux}. 

\section{Thermodynamically-consistent anelastic LES models}

We have obtained filtered budgets of mass, entropy, momentum and energy
that are exact in the turbulent limit, but for the purpose of predicting
filtered quantities, a set of filtered prognostic variables must be
chosen and closure formulae are required that yield turbulent quantities
given the prognostic variables. In this section we aim at identifying
how the various closure relationships must be related in order to
ensure conservation of energy and production of entropy. This problem
is not solved in full generality. Rather, we make a small number of
assumptions from which we can work out an explicit expression for
entropy production that ensures the conservation of total energy,
while leaving as many closures as possible free. 

\subsection{Prognostic equations}

In addition to $q,\mathbf{u}$, we choose to prognose :
\begin{itemize}
\item specific entropy $s$, in order to have access to entropy production
and check its positivity
\item TKE $k$, which requires closures for $\overline{b'w'}$ and $\varepsilon$.
\end{itemize}
Furthermore we neglect non-linearities in the thermodynamic functions,
and assume that thermodynamic relationships apply to filtered thermodynamic
variables. The resulting set of prognostic equations is:
\begin{align}
\rho & =\rho_{ref}\label{eq:rho_prog}\\
\nabla_{t}\rho+\nabla\cdot\rho\mathbf{u} & =0\\
\rho D_{t}q+\nabla\cdot\rho\,\widehat{q'\mathbf{u}'} & =0\label{eq:q_prog}\\
\rho D_{t}s+\nabla\cdot\rho\,\widehat{s'\mathbf{u}'} & =\widehat{\sigma_{s}}\label{eq:s_prog}\\
\rho D_{t}u+\nabla\cdot\rho\widehat{u'\mathbf{u}'}+\rho\nabla_{x}\frac{\pi}{\rho} & =0\\
\rho D_{t}v+\nabla\cdot\rho\widehat{v'\mathbf{u}'}+\rho\nabla_{y}\frac{\pi}{\rho} & =0\\
\rho D_{t}w+\nabla\cdot\rho\widehat{w'\mathbf{u}'}+\rho\nabla_{z}\frac{\pi}{\rho} & =b\\
\rho D_{t}k+\nabla\cdot\widehat{\mathbf{J}_{k}} & =\rho \widehat{P}-\widehat{\varepsilon}\nonumber \\
 & -\rho\left(\widehat{u'\mathbf{u}'}\cdot\nabla u+\widehat{v'\mathbf{u}'}\cdot\nabla v+\widehat{w'\mathbf{u}'}\cdot\nabla w\right)\label{eq:k_prog}
\end{align}
where $\widehat{X}$ designates a closure formula approximating $\overline{X}$ and 
\begin{equation}
\widehat{P} = \widehat{b'w'} \label{TKE_production}
\end{equation}
is the closure for the production of TKE by buoyancy forces.
.

\subsection{Energy-conserving entropy production}

Before specifying more precisely closure assumptions, we derive an
energy budget from (\ref{eq:rho_prog}-\ref{eq:k_prog}) and choose
entropy production $\widehat{\sigma_{s}}$ in such a way that the
energy budget is closed.

The equations of motion (\ref{eq:rho_prog}-\ref{eq:k_prog}) imply
: 
\begin{align}
\rho D_{t}\left(k+\frac{\mathbf{u}\cdot\mathbf{u}}{2}\right)= & \rho\widehat{P}-\widehat{\varepsilon}-\nabla\cdot\left(\rho\mathbf{u}\cdot\widehat{\mathbf{u}'\otimes\mathbf{u}'}+\widehat{\mathbf{J}_{k}}\right)\\
\rho D_{t}h= & T\widehat{\sigma_{s}}+\rho\left(\widehat{s'\mathbf{u}'}\cdot\nabla T+\widehat{q'\mathbf{u}'}\cdot\nabla h_{q}\right)-\nabla\cdot\rho\left(T\widehat{s'\mathbf{u}'}+h_{q}\widehat{q'\mathbf{u}'}\right)\label{eq:prog_enthalpy}
\end{align}
so that, letting $E_{LES}=E+\rho k=\rho(\mathbf{u}\cdot\mathbf{u}/2+k+\Sigma)-p_{ref}$
and using (\ref{eq:hq}): 
\begin{align}
\nabla_{t}E_{LES}+\nabla\cdot\left[(E_{LES}+p_{ref}+\pi)\mathbf{u}+\widehat{\mathbf{J}_{E}}\right]= & \rho\left(\widehat{s'\mathbf{u}'}\cdot\nabla T+\widehat{q'\mathbf{u}'}\cdot\nabla h_{q}\right)+T\widehat{\sigma_{s}}+\rho\widehat{P}-\widehat{\varepsilon}\label{eq:energy_budget}\\
\text{where }\widehat{\mathbf{J}_{E}}\equiv & \rho\left(h^{q}\widehat{q'\mathbf{u}'}+T\widetilde{s'\mathbf{u}'}+\mathbf{u}\cdot\widehat{\mathbf{u}'\otimes\mathbf{u}'}\right)+\widehat{\mathbf{J}_{k}}\label{eq:J_E}
\end{align}
Let $\widehat{\sigma_{s}}=\widehat{\varepsilon}/T+\widehat{\sigma_{mix}}$
with $\widehat{\sigma_{mix}}$ a source of entropy due to turbulent
mixing, which adds to the source due to viscous dissipation. Then
:
\begin{equation}
\nabla_{t}E_{LES}+\nabla\cdot\left[(E_{LES}+p_{ref}+\pi)\mathbf{u}+\widehat{\mathbf{J}_{E}}\right]=\rho\left(\widehat{s'\mathbf{u}'}\cdot\nabla T+\widehat{q'\mathbf{u}'}\cdot\nabla\mu+\widehat{P}\right)+T\widehat{\sigma_{mix}}
\end{equation}
A \emph{sufficient condition} to close the energy budget is then that:
\begin{equation}
T\widehat{\sigma_{mix}}=-\rho\left(\widehat{s'\mathbf{u}'}\cdot\nabla T+\widehat{q'\mathbf{u}'}\cdot\nabla\mu+\widehat{P}\right)\label{eq:second_law-1}
\end{equation}
A set of closures such that $\widehat{\varepsilon}\ge0$ and $T\widehat{\sigma_{mix}}\ge0$
is then consistent with the second law.

It should be emphasized that (\ref{eq:second_law-1}) may not be the
only way to close the energy budget. Indeed the identification of
an energy flux and an entropy source is not unique: it may be possible
to adopt a different expression for $\mathbf{J}_{E}$, which would
change (\ref{eq:second_law-1}). The identification made here uses
the heuristics that the entropy source and the energy flux can be
expressed locally from the other fluxes. The validity of this assumption
certainly depends on other assumptions, especially those made to close
the turbulent fluxes.

Finallly, comparing expression (\ref{eq:J_E}) of the modelled total
energy flux to its exact expression (\ref{eq:total_energy_flux})
leads to the conclusion that the closures for the turbulent sensible
heat flux and the reduced turbulent entropy flux are not independent,
but related by :
\begin{equation}
\widehat{\mathbf{J}_{H}}=\rho T\widetilde{s'\mathbf{u}'}\label{eq:closure_JH}
\end{equation}
i.e. relationship (\ref{eq:reduced_fluxes}) relating molecular fluxes
also applies to turbulent fluxes in this setting.

\subsection{Closure assumptions for a positive entropy production}

Since $\hat{\varepsilon}$ is the output of a closure, it is quite
straightforward to demand that it be positive. Conversely, since the
condition $T\widehat{\sigma_{mix}}\ge0$ depends on the closures for
$\widehat{s'\mathbf{u}'}$, $\widehat{q'\mathbf{u}'}$ and $\widehat{P}$,
it is less obvious how restrictive it is. In order to simplify this
condition, we shall assume that $\widehat{P}$ is not an independent
closure, and rather can be deduced from $\widehat{s'\mathbf{u}'},\widehat{q'\mathbf{u}'}$
as:
\begin{equation}
\widehat{P}=b_{s}\widehat{s'w'}+b_{q}\widehat{q'w'}\label{eq:assumption_bw}
\end{equation}
This assumption is consistent with the heuristics that the perturbation
$X'$ for quantities $X=q,s$ that obey a pure transport equation
under adiabatic conditions can be estimated as $X'\simeq-z'\nabla_{z}X$
with $z'$ a vertical displacement above a fluid-parcel dependent
reference level such as its level of neutral buoyancy. We note for
future reference that, since $b=-\Sigma_{z}$:
\begin{equation}
b_{s}=-h_{sz},\qquad b_{q}=-h_{qz}\label{eq:buoyancy_derivatives}
\end{equation}

Now 
\[
\nabla T=h_{sz}\nabla z+h_{ss}\nabla s+h_{sq}\nabla q,\qquad\nabla h_{q}=h_{qz}\nabla z+h_{qs}\nabla s+h_{qq}\nabla q
\]
so that :

\begin{align}
T\widehat{\sigma_{mix}}= & -\rho h_{ss}\widehat{s'\mathbf{u}'}\cdot\nabla s-\rho h_{sq}\left(\widehat{s'\mathbf{u}'}\cdot\nabla q+\widehat{q'\mathbf{u}'}\cdot\nabla s\right)-\rho h_{qq}\widehat{q'\mathbf{u}'}\cdot\nabla q\label{eq:entropy_production}
\end{align}
One recognizes in (\ref{eq:entropy_production}) the symmetric bilinear
form $\sigma_{iso}$ expressing the entropy produced by the isobaric
mixing of two fluid parcels (see subsection \ref{subsec:Entropy-production-turbulent}):
\begin{equation}
\widehat{\sigma_{mix}}=-\rho\,\sigma_{iso}\left[(\widetilde{s'\mathbf{u}'},\widehat{q'\mathbf{u}'}),\,(\tilde{\nabla}s,\nabla q)\right].\label{eq:turbulent_entropy_prod-1}
\end{equation}

It is interesting to compare expression (\ref{eq:turbulent_entropy_prod-1})
to the fine-grain entropy source $\sigma_{mix}$ (\ref{eq:entropy_source}).
The latter is a bilinear expression of the reduced fluxes $\mathbf{j}'_{s},\,\mathbf{j}_{q}$
and the thermodynamic forces $\nabla T$, $\nabla\mu-\mu^{T}\nabla T$.
The standard approach of non-equilibrium thermodynamics is to postulate
a linear relationship between fluxes and thermodynamic forces, whose
coefficients must guarantee $\sigma_{mix}\ge0$ \citep{groot_non-equilibrium_1962}.
The same approach applied to (\ref{eq:turbulent_entropy_prod-1})
results in a linear relationship between the reduced turbulent fluxes
and the ``turbulent thermodynamic forces'' $\tilde{\nabla}s,\,\nabla q$,
which is precisely a down-gradient closure. \\

We therefore assume now that $\widetilde{s'\mathbf{u}'}$, $\widehat{q'\mathbf{u}'}$
are related linearly to $\tilde{\nabla}s,\,\nabla q$. Leaving the
consideration of anisotropic three-dimensional diffusion for future
work:
\begin{align}
\widehat{q'\mathbf{u}'} & =-K_{qs}\widetilde{\nabla}s-K_{qq}\nabla q\label{eq:closure_q}\\
\widetilde{s'\mathbf{u}'} & =-K_{ss}\widetilde{\nabla}s-K_{sq}\nabla q\label{eq:closure_s}
\end{align}
with $K_{xy}$ flow-dependent coefficients. It should be noted that
this set of closure formulae is equivalent to another set using the
full entropy flux and gradient. However the coefficients that appear
in this other set depend on reference enthalpies and entropies in
a non-trivial way, while coefficients $K_{xy}$ are invariant by design.

(\ref{eq:closure_q}-\ref{eq:closure_s}) allow for cross-diffusion.
A more common assumption is to let $K_{qs}=K_{sq}=0$. In this case
(\ref{eq:turbulent_entropy_prod-1}) simplifies to:
\begin{equation}
T\widehat{\sigma_{mix}}=\rho\left(K_{ss}h_{ss}\left(\tilde{\nabla}s\right)^{2}+g^{qq}K_{qq}\left(\nabla q\right)^{2}\right) \label{eq:closure_entropy_production}
\end{equation}
i.e. down-gradient turbulent diffusion, including with differential
turbulent diffusion of heat and matter, is consistent with the second
law as soon as $K_{ss}\ge0$ and $K_{qq}\ge0$. This of course includes
the common assumption of a single mixing coefficient $K_{ss}=K_{qq}\ge0$. 

If cross-diffusion is allowed, a sufficient condition for $\widehat{\sigma_{mix}}\ge0$
is that the quadratic form :
\[
K_{ss}h_{ss}\widetilde{\text{d}s}^{2}+\left(K_{sq}h_{ss}+K_{qs}g^{qq}\right)\widetilde{\text{d}s}\text{d}q+K_{qq}g^{qq}\text{d}q^{2}
\]
be positive definite. With $K_{ss}\ge0$, $K_{qq}\ge0$, this condition
is equivalent to :
\begin{align}
\left(K_{sq}h_{ss}+K_{qs}g^{qq}\right)^{2} & \le4K_{ss}K_{qq}h_{ss}g^{qq}\label{eq:cross_diffusion}
\end{align}
which sets an upper bound to the magnitude of cross-diffusion.

\subsection{Boundary conditions}

Evolution equations (\ref{eq:rho_prog}-\ref{eq:k_prog}) must be
complemented by boundary conditions, which typically relate turbulent
fluxes to boundary values. For the sake of invariance, these boundary
conditions should involve invariant fluxes and boundary quantities.
The sensible heat flux $\widehat{\mathbf{J}_{H}}$ (or equivalently
the reduced entropy flux) and compositional flux $\widehat{q'\mathbf{u}'}$
are natural candidates. For instance, at the lower, solid boundary
of an atmospheric domain, one could use, as in \cite{pressel_large-eddy_2015} :
\begin{equation}
\widehat{\mathbf{J}_{H}}\cdot\mathbf{n}=\rho C_{h}U_b\,C_{p}(T_{b}-T_{surf}),\qquad\widehat{q'\mathbf{u}'}\cdot\mathbf{n}=C_{q}U_b\,\left(q_{b}-q_{sat}(p_{surf},\,T_{surf})\right) \label{eq:bulk_formulae}    
\end{equation}
with $C_{h}$ (resp. $C_{q}$ ) an exchange coefficient for heat (resp.
moisture), $U_{b}, T_{b},\,q_{b}$ atmospheric values of 
$\left\Vert \mathbf{u} \right\Vert, \, T, \, q$
at a certain height above ground, $T_{surf},\,p_{surf}$ the value of $T,\,p$ at the soil surface
and $q_{sat}(p,\,T)$ specific humidity at saturation.

\section{Thermodynamic invariance of the LES models}

\subsection{Temperature equation}

At this point the LES model (\ref{eq:rho_prog}-\ref{eq:k_prog})
has been completed by explicit closures, at least regarding those
that matter for energetics and thermodynamics. It is thus possible
to examine its thermodynamic invariance properties. To this end it
is useful to form an evolution equation for temperature. Using (\ref{eq:assumption_bw},
\ref{eq:entropy_production}), the evolution equation for enthalpy
(\ref{eq:prog_enthalpy}) becomes :
\begin{equation}
\rho\nabla_{t}h+\nabla\cdot\left(\rho\left(T\widehat{s'\mathbf{u}'}+h_{q}\widehat{q'\mathbf{u}'}\right)\right)=\widehat{\varepsilon}-\rho\widehat{P}\label{eq:enthalpy_equation}
\end{equation}
Now using (\ref{eq:q_prog}), $h_{q}=g^{q}=h^{q}-Ts^{q}$ and $\nabla_{t}h=\rho c_{p}\nabla_{t}T+\rho h^{q}\nabla_{t}q$
: 
\begin{align}
\rho c_{p}\nabla_{t}T+\rho\widehat{q'\mathbf{u}'}\cdot\nabla h^{q}+\nabla\cdot\widehat{\mathbf{J}_{H}} & =\widehat{\varepsilon}-\rho\widehat{P}\label{eq:temperature_equation}
\end{align}
Our closures have been designed precisely to ensure that the turbulent
fluxes $\widehat{q'\mathbf{u}'}$ and $\widehat{\mathbf{J}_{H}}$
are invariant under the changes considered by \cite{dubos_thermodynamic_2024},
i.e. by adding a constant to partial enthalpies and entropies. Furthermore
definition (\ref{eq:assumption_bw}) relates $\widehat{P}$ to
$\tilde{\nabla}_{z}s$ and $\nabla_{z}q$ with precisely the coefficients
that relate $N^{2}$ to $\widetilde{\nabla}s$ and $\nabla q$ in
expression (\ref{eq:BruntVaisala}): 
\[
\widehat{P}=b_{s}\widetilde{s'w'}+\left(b_{q}+b_{s}s^{q}\right)\widehat{q'w'},\quad N^{2}=b_{s}\tilde{\nabla}_{z}s+\left(b_{q}+b_{s}s^{q}\right)\nabla_{z}q.
\]
Since $N^{2},\,\tilde{\nabla}_{z}s,\,\nabla_{z}q,\,b_{s}=\Sigma_{zs}=T_{z}$
are invariant, $b_{q}+b_{s}s^{q}$ hence $\widehat{P}$ and finally
$\nabla_{t}T$ are invariant.

\subsection{A stronger invariance}

Furthermore, one may ask whether the filtered model is invariant with
respect to changes of the Gibbs function that are not allowed for
the unfiltered equations. Especially, let us consider the effect of
adding a term $-T\delta s(q)$ to $g(z,T,q)$:
\begin{equation}
g\rightarrow g-T\delta s(q)\;\Leftrightarrow\;s\rightarrow s+\delta s(q)\label{eq:invariance}
\end{equation}
with $\delta s(q)$ an arbitrary, possibly non-linear function of
$q$. (\ref{eq:invariance}) does not affect enthalpy $h=g-Tg^{T}$,
heat capacity $c_{p}=-Tg^{TT}$, equation of state $\upsilon=g^{p}$
or stability $N^{2}$ (\ref{eq:BruntVaisala}). (\ref{eq:invariance})
also leaves adiabatic curves $(s,q)=cst$ unchanged in $(p,T,q)$
space. Since adiabatic curves are the same, buoyancy $b=-\Sigma_{z}$
is also unchanged. However (\ref{eq:invariance}) does affect $g^{qq}$
hence $\nabla\mu-\mu^{T}\nabla T=g^{pq}\nabla p+g^{qq}\nabla q$.
Physically this is because the ``true'' $s(p,T,q)$ includes a mixing
entropy term $s_{mix}(q)$ whose expression is physically meaningful
and not arbitrary. Molecular diffusive fluxes, which are linear in
$\nabla\mu-\mu^{T}\nabla T$ and $\nabla T$ (see section 2), are
thus affected, so that unfiltered anelastic equations are not invariant
with respect to (\ref{eq:invariance}). 

Isobaric entropy production $\sigma_{iso}$ is sensitive to (\ref{eq:invariance}).
As a result, in our LES model, the coarse-grain entropy production
$\widehat{\sigma_{mix}}$ \emph{is} sensitive to (\ref{eq:invariance}).
However: 
\begin{equation}
\delta\nabla s=\nabla\delta s=\frac{\text{d}\delta s}{\text{d}q}\nabla q,\qquad\delta s^{q}=\frac{\text{d}\delta s}{\text{d}q},\qquad\delta\widetilde{\nabla s}=\delta\nabla s-\delta s^{q}\nabla q=0,
\end{equation}
so that down-gradient turbulent fluxes (\ref{eq:closure_q}-\ref{eq:closure_s})
and the temperature equation (\ref{eq:temperature_equation}) are
not affected. Overall, (\ref{eq:invariance}) has no effect on the
fields $(\mathbf{u},T,q,k)$ predicted by the LES model (\ref{eq:rho_prog}-\ref{eq:k_prog},
\ref{eq:assumption_bw}, \ref{eq:closure_q}, \ref{eq:closure_s}),
which therefore is invariant under a wider class of transformations
than the unfiltered model (\ref{eq:AN_EOS}-\ref{eq:AN_momentum}).

\subsection{Prognosing a conservative variable}

It may be desirable to prognose, rather than entropy itself, a conservative
variable $\theta=\theta(s,q)$ such as potential temperature $T(p_{0},s,q)$
or potential enthalpy $h(p_{0},s,q)$, even though such variables
depend on a purely arbitrary reference pressure $p_{0}$. Upon performing
this change of variables, one obtains the prognostic equation: 
\begin{align}
\rho\nabla_{t}\theta+\nabla_{z}\left(\rho\widehat{\theta'\mathbf{u}'}\right) & =\frac{\widehat{\varepsilon}}{h_{\theta}}+\widehat{\sigma_{\theta}^{mix}}\label{eq:prog_theta}
\end{align}
and the closures:
\begin{align}
h_{\theta}\widehat{\sigma_{\theta}^{mix}} & =-\rho h_{\theta\theta}\widehat{\theta'\mathbf{u}'}\cdot\nabla\theta-\rho h_{\theta q}\left(\widehat{\theta'\mathbf{u}'}\cdot\nabla q+\widehat{q'\mathbf{u}'}\cdot\nabla\theta\right)-\rho h_{qq}\widehat{q'\mathbf{u}'}\cdot\nabla q\label{eq:closure_sigma_theta}\\
\widehat{P} & =h_{z\theta}\widehat{\theta'w'}+h_{zq}\widehat{q'w'}\label{eq:closure_bw_theta}
\end{align}
where $h_{\theta},\,h_{\theta\theta},\,h_{\theta q},\,h_{z\theta},\,h_{zq},\,h_{qq}$
are now derivatives of $h(z,\theta,q)$. Closures (\ref{eq:closure_sigma_theta},\ref{eq:closure_bw_theta})
require only the knowledge of the expression $h(z,\theta,q)$. Also,
adding the divergence $\nabla\cdot\mathbf{F}_{rad}$ of a radiative
flux to the enthalpy equation is equivalent to adding $(1/h_{\theta})\nabla\cdot\mathbf{F}_{rad}$
to (\ref{eq:prog_theta}), which is also possible from the knowledge
of $h(z,\theta,q)$ only. 

The source term $\widehat{\sigma_{\theta}^{mix}}$ ensures conservation
of total energy. In general $h$ is not convex w.r.t. $(\theta,q)$
so that $\widehat{\sigma_{\theta}^{mix}}$ is not positive definite.
This is not problematic since $\widehat{\sigma_{\theta}^{mix}}$
is not directly related to entropy production. 

In order to compute $\widehat{\theta'w'}$ and $\widehat{q'w'}$ from
$\nabla\theta,\,\nabla q$, one writes:
\begin{align}
\widehat{\theta'w'} & =\theta_{s}\widetilde{s'w'}+\left(\theta_{q}+\theta_{s}s^{q}\right)\widehat{q'w'}\label{eq:closure_theta_flux}\\
\nabla\theta= & \theta_{s}\widetilde{\nabla s}+\left(\theta_{q}+\theta_{s}s^{q}\right)\nabla q
\end{align}
where $\theta_{s},\,\theta_{q}$ are the derivatives of $\theta(s,q)$,
and uses (\ref{eq:closure_sigma_theta},\ref{eq:closure_bw_theta}).
Thus computing $\widehat{\theta'w'}$, $\widehat{q'w'}$ from $\nabla\theta,\,\nabla q$
requires that the functions $s^{q}$ and $\theta(s,q)$ be known,
in general. However there is a special case: if $K_{qs}=K_{sq}=0$
and $K_{ss}=K_{qq}$, (\ref{eq:closure_q},\ref{eq:closure_s}) simplify
to: 
\begin{equation}
\widehat{s'w'}=-K_{ss}\nabla s,\qquad\widehat{q'w'}=-K_{ss}\nabla q\qquad\Rightarrow\qquad\widehat{\theta'w'}=-K_{ss}\nabla\theta.
\end{equation}
so that $\theta,\,q$ can be prognosed while ignoring the precise
relationship $\theta(s,q)$. Notice that relating the turbulent sensible
heat flux (\ref{eq:closure_JH}) to $\nabla\theta,\,\nabla q$ still
involves $\theta_{s},\,\theta_{q},\,s^{q}$. Thus the knowledge of
$\theta(s,q)$ is not necessary in the fluid interior, but remains
needed for coupling to boundary heat fluxes.
\subsection{Indifference to $\theta(s,q)$}
We can ask, conversely, which down-gradient closures for the turbulent fluxes are insensitive to the relationship $\theta(s,q)$. A safe way to construct
such closures is to base them on quantities that are themselves insensitive. In terms of fluxes, natural candidates are  
$\widehat{P}$ (which is not strictly speaking a flux since $b(z,s,q)$ depends, in general, on $z$) and $\widehat{q'w'}$. The corresponding gradients are $N^2$
(\ref{eq:BruntVaisala}) and $\nabla q$. The most general closure would assume a linear relationship between $(\widehat{P}, \widehat{q'w'})$ and  
$(N^2, \nabla q)$, possibly with cross terms and different diagonal coefficients. 

However, to establish the consistency of such a closure with the second law, one must convert it to the form (\ref{eq:closure_s}, \ref{eq:closure_q}), which in general requires the knowledge of the relationship $\theta(s,q)$. Again, the apparently only exception is when one assumes a single mixing coefficient. 
Thus, it seems that closures that are both indifferent to $\theta(s,q)$ and unconditionnally consistent with the second law are those with a single, positive mixing coefficient. 

\section{Compressible LES models for gravity-dominated flows}
Instead of resorting to the anelastic approximation, it is possible to model low-Mach number flows with compressible equations (e.g. \citealt{romps_dry-entropy_2008, thuburn_potential-based_2025}). 
It is therefore desirable to adapt the family of tubulent closures considered in section 4 to compressible models, which is the aim of this section. This adaptation is heuristic, in the
sense that it is not based on a derivation of exact budgets similar
to those of section 3. Instead, the adiabatic part of the model is
replaced by the compressible Euler equations and the turbulent closure
$\widehat{P}$ is slightly modified. 

In a compressible model,
there is no notion of a reference pressure profile $p_{ref}(z)$.
Thermodynamic functions become again functions of $(p,\,s,\,q)$ and
the equation of state is $\rho^{-1}=\upsilon=h_{p}$. Replacing the
adiabatic part of (\ref{eq:rho_prog}-\ref{eq:k_prog}) by Euler equations
yields:
\begin{align}
\frac{1}{\rho} & =h_{p}\label{eq:rho_prog-1}\\
\nabla_{t}\rho+\nabla\cdot\rho\mathbf{u} & =0\\
\rho D_{t}q+\nabla\cdot\rho\,\widehat{q'\mathbf{u}'} & =0\label{eq:q_prog-1}\\
\rho D_{t}s+\nabla\cdot\rho\,\widehat{s'\mathbf{u}'} & =\frac{\widehat{\varepsilon}}{T}+\widehat{\sigma_{mix}}\label{eq:s_prog-1}\\
\rho D_{t}u+\nabla\cdot\rho\widehat{u'\mathbf{u}'}+\nabla_{x}p & =0\\
\rho D_{t}v+\nabla\cdot\rho\widehat{v'\mathbf{u}'}+\nabla_{y}p & =0\\
\rho D_{t}w+\nabla\cdot\rho\widehat{w'\mathbf{u}'}+\nabla_{z}p & =-\rho\,g_{grav}\\
\rho D_{t}k+\nabla\cdot\widehat{\mathbf{J}_{k}} & =\rho \widehat{P}-\widehat{\varepsilon}\nonumber \\
 & -\rho\left(\widehat{u'\mathbf{u}'}\cdot\nabla u+\widehat{v'\mathbf{u}'}\cdot\nabla v+\widehat{w'\mathbf{u}'}\cdot\nabla w\right)\label{eq:k_prog-1}
\end{align}
With $\widehat{\mathbf{J}_{E}}$ given by (\ref{eq:J_E}), total energy $E_{LES}^{FC}=\rho(\mathbf{u}\cdot\mathbf{u}/2+k+\Phi+h)-p$
is conserved:
\begin{equation}
\nabla_{t}E_{LES}^{FC}+\nabla\cdot\left[(E_{LES}^{FC}+p)\mathbf{u}+\widehat{\mathbf{J}_{E}}\right] =0
\end{equation}
 provided 
(\ref{eq:second_law-1}) holds. Expanding the gradients of $T=h_s$ and $\mu=h_q$ in terms of $\nabla p$, $\nabla s$, $\nabla q$ in (\ref{eq:second_law-1}) leads to:

\begin{equation}
\frac{T \widehat{\sigma_{mix}}}{\rho} = 
T \sigma_{iso}\left[(\widetilde{s'\mathbf{u}'},\widehat{q'\mathbf{u}'}),\,(\tilde{\nabla}s,\nabla q)\right]
-\left(\upsilon_{s}\,\widehat{s'\mathbf{u}'} + \upsilon_{q}\,\widehat{q'\mathbf{u}'}\right)\cdot\nabla p
-\widehat{P}.\label{eq:turbulent_entropy_prod-2}
\end{equation}

In the inertial range of isotropic incompressible turbulence, pressure spectra are steeper than spectra for transported scalars. At a scale $l$, pressure fluctuations are expected to scale as $l^{2/3}$ while those of a transported scalar are expected to scale as $l^{1/3}$ \citep{lighthill_sound_1952}. Thus, in sufficiently developed low-Mach-number turbulence, temperature gradients at dissipative scales are dominated by the contribution due to gradients of entropy and composition. As discussed in section 2, entropy production is then effectively isobaric. In the filtered model, the entropy source can therefore only depend on information present in $\sigma_{iso}$, hence we ask that (\ref{eq:turbulent_entropy_prod-1}) holds as in the anelastic case. Especially $\widehat{\sigma_{mix}}\ge0$ under the same conditions
as in section 4.

(\ref{eq:turbulent_entropy_prod-2}) with (\ref{eq:turbulent_entropy_prod-1})
leads to he closure:
\begin{equation}
\widehat{P}=-\left(\upsilon_q\,\widehat{s'\mathbf{u}'}+\upsilon_{q}\widehat{q'\mathbf{u}'}\right)\cdot\nabla p\label{eq:assumption_bw_FC}
\end{equation}
The rationale behind (\ref{eq:assumption_bw_FC}) is as follows. According to (\ref{eq:turbulent_entropy_prod-2}), if turbulent fluxes were along isobars, we could assume isobaric entropy production (\ref{eq:turbulent_entropy_prod-1}) and close the energy budget with $\widehat{P}=0$, i.e. without consuming/releasing TKE. In fact turbulent fluxes are not along isobars, and $\widehat{P}$ is precisely the amount of TKE that needs to be consumed or released to close the energy budget.
Since (\ref{eq:assumption_bw_FC}) reduces to (\ref{eq:assumption_bw}) as  $p \rightarrow p_{ref}(z)$, both are instances of the same closure for $\widehat{P}$, (\ref{eq:assumption_bw_FC}) being more general while (\ref{eq:assumption_bw}) has
a more straightforward physical interpretation. Interestingly, with (\ref{eq:assumption_bw_FC}), $\widehat{P}$
depends on the full fluxes $\widehat{s'\mathbf{u}'},\,\widehat{q'\mathbf{u}'}$
and not only on the vertical fluxes $\widehat{s'w'},\,\widehat{q'w'}$.
In gravity-dominated flows, the pressure gradient $\nabla p$ is dominantly
vertical, but it is only in the anelastic limit $p\rightarrow p_{ref}$
that it becomes purely vertical.

\section{Discussion}

\subsection{Source term and indifference to reference values}

The source terms $\widehat{\varepsilon}/h_{\theta}$ and $\widehat{\sigma_{\theta}^{mix}}$
of (\ref{eq:prog_theta}) are often neglected in practice in numerical
models. In fact, $\widehat{\sigma_{\theta}^{mix}}$ vanishes whenever
$h(z,\theta,q)$ is linear with respect to $\theta,q$. This happens
for dry air modelled as an ideal perfect gas and for water modelled
with a linearized equation of state. Such degenerate situations simplify
(\ref{eq:prog_theta}), but it should be emphasized that they are
the exception rather than the rule. In general, neglecting these source
terms implicitly modifies the value of the entropy source in (\ref{eq:s_prog}).
This potentially breaks the conservation of total energy, the positivity
of entropy production and invariance of the prognostic equations with
respect to reference pressure, enthalpies and/or entropies \citep{dubos_thermodynamic_2024}.
Starting from a given distribution of temperature and moisture, different
choices of reference values may then lead to different model predictions.
Furthermore, any energetic analysis of the resulting system will present
internal inconsistencies. Thus, while approximating or neglecting
$\widehat{\varepsilon}/h_{\theta}$ and/or $\widehat{\sigma_{\theta}^{mix}}$
may be justified quantitatively when producing approximate numerical
solutions, these terms must be retained in the prognostic equations
for the purpose of analyzing their energetics or invariance.

\subsection{Turbulent heat fluxes from colder to warmer fluid}

In this subsection, the question of the consistency of turbulent heat
exchange from colder fluid to warmer fluid with the second law of
thermodynamics is briefly discussed. For the sake of simplicity, the
fluid is assumed to be dry air modelled as an ideal perfect gas with
constant heat capacity $C_{p}$ and $\kappa=R/C_{p}$ and the anelastic
system is defined with respect to an adiabatic reference profile $\theta=\theta_{ref}$.
In this context,
\begin{equation}
s=C_{p}\log\frac{\theta}{T_{00}},\qquad h=h_{00}+C_{p}\theta\left(\frac{p}{p_{0}}\right)^{\kappa},\qquad p_{ref}=p_{s}\left(1-\frac{\Phi}{C_{p}T_{s}}\right)^{1/\kappa}
\end{equation}
\begin{equation}
\Sigma(z,\theta)=gz+h_{00}+C_{p}\theta\left(\frac{p_{s}}{p_{0}}\right)^{\kappa}\left(1-\frac{\Phi}{C_{p}T_{s}}\right),\qquad b=-\Sigma_{z}=g\left(\frac{\theta}{\theta_{ref}}-1\right)
\end{equation}
with $p_{0},\,h_{00},\,T_{00}$ irrelevant constants and $T_{s}=\theta_{ref}(p_{s}/p_{0})^{\kappa},p_{s}$
the temperature and pressure at $z=0$. Letting $p_{0}=p_{s}$ for
simplicity and using $h_{ss}=T/C_{p}$:
\begin{equation}
T=\Sigma_{s}=\theta\left(1-\frac{\Phi}{C_{p}T_{s}}\right),\qquad-\nabla_{z}T=\left[\frac{g}{C_{p}}\,\frac{\theta}{T_{s}}-\left(1-\frac{\Phi}{C_{p}T_{s}}\right)\nabla_{z}\theta\right]\label{eq:lapse_rate}
\end{equation}
\begin{equation}
\widehat{P}=\frac{g}{C_{p}}\widehat{s'w'}=\frac{g}{T_s}\widehat{\theta'w'},\qquad\mathbf{J}_{H}=\rho T\widehat{s'\mathbf{u}'}=\rho C_{p}\widehat{\theta'\mathbf{u}'},\qquad\widehat{\sigma_{mix}}=\rho K_{ss}\frac{\left\Vert \nabla s\right\Vert ^{2}}{C_{p}}
\end{equation}
According to (\ref{eq:lapse_rate}), one may have in a weakly stratified
environment $\nabla_{z}\theta>0$ and $-\nabla_{z}T\ge0$. The turbulent
heat flux $\mathbf{J}_{H} \sim \widehat{s'\mathbf{u}'} \sim \widehat{\theta'\mathbf{u}'} \sim-\nabla\theta$
is then downwards, from colder air to warmer air. While this may appear
to contradict the second law, it does not: indeed $\widehat{\sigma_{mix}}\ge0$
as soon as $K_{ss}\ge0$. 

\subsection{Assumed-TKE models}

Simpler models are obtained by omitting $k$ as a prognostic field and assuming an instantaneous
balance between production and destruction of TKE:
\begin{equation}
\widehat{\varepsilon}=\rho\widehat{b'w}+P_{shear}\label{eq:zero_TKE}
\end{equation}
so that the enthalpy and entropy equations become :
\begin{align}
\rho\nabla_{t}h+\nabla\cdot\left(\rho\left(\widehat{h_{q}}\widehat{q'\mathbf{u}'}+T\widehat{s'}\mathbf{u}'\right)\right) & =P_{shear}\\
\rho TD_{t}s+T\nabla\cdot\rho\,\widehat{s'\mathbf{u}'}= & \rho\widehat{b'w}+P_{shear}+T\widehat{\sigma_{s}^{mix}}
\end{align}
This approach is typical of coarse-resolution ocean and atmospheric
general circulation models (GCMs). Additionnally, GCMs often neglect
the entropy source term which has a numerically small impact. However
this omission hides crucial information about the energetics and thermodynamics
of the closures, since no term of the equations reflects the second
law or the fact that turbulent fluxes produce or consume TKE. For
instance, shear production of TKE $P_{shear}$ appears as a sink term
in the resolved kinetic energy equation. It is thus tempting to conflate
$P_{shear}$ with $\widehat{\varepsilon}$ \citep{gassmann_entropy_2018}.
In fact (\ref{eq:zero_TKE}) implies that $\widehat{\varepsilon}$
may be larger or smaller than $\rho P_{shear}$ depending on whether
turbulent mixing releases or consumes TKE. Even when the entropy production
term is included, it requires some physical intuition to infer that
closures are constrained only by (\ref{eq:zero_TKE}) and the requirement
that $\widehat{\varepsilon}\ge0$ \citep{akmaev_energetics_2008}.
The present approach provides a solid and
unambiguous basis to establish such thermodynamic constraints
on down-gradient closures.

\subsection{Invariance of filtered models}

In section 5, the filtered model is found to be invariant under a wider class of
transformations than the unfiltered model. Also, in the special case
of a single turbulent diffusion, incomplete thermodynamic information
is sufficient to prognose a conservative variable, i.e. knowing the
expression $h(p,\theta,q)$ is sufficient. This property is close
to an even stronger property of the adiabatic dynamics, for which
knowledge of the equation of state $\upsilon(p,\theta,q)=h_{p}$ is
sufficient. This observation suggests to place quantities involved in the model
on a scale going from purely reversible/mechanical to irreversible/thermal.
For instance $\widehat{q'w'}$ and $\widehat{P}$ are adiabatic/mechanical quantities while the
turbulent sensible heat flux $\mathbf{J}_{H}$ or the entropy source
are intrinsically ``thermal'' quantities.

Geophysical turbulence at various scales is often ``inviscidly-driven'', i.e. the result of inviscid instabilities such as shear instability, static instability, symmetric instability or baroclinic instability. Furthemore what matters for inviscid dynamics are the adiabats $\theta=cst$ in $(p,\upsilon)$ space for a given $q$, but not the value $\theta(s,q)$ on these adiabats. Thus the turbulent flux $\widehat{\theta'w'}$ should be closed the same way for any $\theta(s,q)$ when turbulence is ``inviscidly-driven''. As discussed in 5.4, having equal mixing coefficients for $\theta$ and $q$ is apparently the only way to satisfy this constraint.

Conversely, one could expect differing mixing coefficients and/or cross-diffusion when the mechanism generating turbulence is not inviscid, e.g. doubly-diffusive convection \citep{middleton_general_2020}. 
Since the energetics of isobaric
mixing involve information present in $h$ but not in the equation
of state alone, it also seems inevitable that computing the source
term for $\theta$ requires knowledge of the full enthalpy function
$h(p,\theta,q)$. Finally, it makes sense that the evaluation of intrinsically
thermal quantities such as the entropy source and the turbulent sensible
heat flux also requires to connect $\theta$ to entropy.

Hopefully a more thorough examination of the above hypotheses can lead to set various expectations on the information
needed in the closures. In the future, such
expectations could constitute a guiding principle to identify admissible
closures or to design new ones.

\subsection{Summary and future work}

Exact filtered budgets for multicomponent anelastic turbulence have
been obtained. Noteworthy results are the enthalpy equation (\ref{eq:enthalpy_budget})
and the invariant expression (\ref{eq:sensible_heat_flux}) for the
turbulent sensible heat flux, valid for arbitrary thermodynamics.
A family of turbulent closures consistent with the first and second
laws of thermodynamics has been identified. The equations of motion
are (\ref{eq:rho_prog}-\ref{eq:k_prog}) or their compressible variant
(section 6). Closures for $\widehat{u'\mathbf{u}'}$, $\widehat{v'\mathbf{u}'}$,
$\widehat{w'\mathbf{u}'}$, $\widehat{\mathbf{J}_{k}}$, $\widehat{\varepsilon}$
are left unspecified: the first and second law do not constrain them
beyond the condition $\widehat{\varepsilon}\ge0$. Closure (\ref{eq:closure_JH})
for the turbulent sensible heat flux derives from the closure for
$\widetilde{s'\mathbf{u}'}$ . The closure for buoyant TKE production
$\overline{w'b'}$ derives from the closures for the fluxes $\widetilde{s'\mathbf{u}'}$
and $\overline{q'\mathbf{u}'}$. Interestingly, in an anelastic model,
(\ref{eq:assumption_bw}) uses their vertical component $\widetilde{s'w'}$
and $\overline{q'w'}$ as is customary while in a fully compressible
model, (\ref{eq:assumption_bw_FC}) uses their projection along
the pressure gradient. These closures being defined, the entropy source
due to turbulent mixing is given by (\ref{eq:turbulent_entropy_prod-1})
and is closely related to the entropy production due to isobaric mixing
(\ref{eq:sigma_iso}).

Assuming down-gradient turbulent fluxes (\ref{eq:closure_q}-\ref{eq:closure_s}),
the entropy source becomes a quadratic function of $\widetilde{\nabla s},\,\nabla q$.
In the absence of cross-diffusion, the positivity of turbulent mixing coefficients
is sufficient to ensure a positive entropy source. In the presence
of cross-diffusion, positivity is subject to condition (\ref{eq:cross_diffusion}).
With or without cross-diffusion, the LES models are thermodynamically invariant, and in fact 
they are invariant with respect to the wider
class (\ref{eq:invariance}) of thermodynamic transformations.
It is straightforward to prognose a conservative variable $\theta(s,q)$
instead of entropy. In the special case of a single turbulent mixing
coefficient, $\theta$ can be prognosed while ignoring is relationship
to entropy, except for the boundary heat fluxes. \\

The present work has a number of limitations, and there are several
ways in which it could be extended in the future. Unlike
many LES models aimed at studying clouds and convection \citep{pressel_large-eddy_2015},
the present framework omits hydrometeors moving at a speed that differs
from the fluid velocity. To overcome this shortcoming in a consistent manner,
one may need to start from a non-averaged model that represents each class of hydrometeor as a distinct fluid, with its own density, velocity and temperature, and spells
out budgets of momentum, energy, and entropy for each fluid, with
a dynamically and thermodynamically consistent treatment of the exchanges
of heat, momentum and mass between fluids. It may also be advantageous to let averages $\overline X$ be weighted by the dry air mass instead of the moist air mass \citep{romps_dry-entropy_2008}. This is left for future work.

Large-scale ocean models, due to the disparate processes acting in
the vertical and in the ``horizontal'' directions, use anisotropic
turbulent diffusion, represented as a $3\times3$ tensor rather than
a scalar \citep{redi_oceanic_1982}. This tensor may be non-symmetric
in order to represent the action of baroclinic eddies \citep{griffies_gentmcwilliams_1998}.
It would be interesting, and relatively straightforward to extend
the present work to this situation.

Total potential energy can be split into background potential energy
BPE and available potential energy \citep{tailleux_simple_2024}. The
buoyant term $\overline{b'w'}$ is usually interpreted as representing
a conversion between TKE and APE \citep{peltier_mixing_2003}, APE
being converted into BPE by conductive and diffusive fluxes. The distinction
between APE and BPE is not explicit in the present LES model. In fact
$\overline{b'w'}$ may directly feed BPE, for example if the resolved
flow varies only in the vertical direction. Turbulence models prognosing,
in addition to TKE, a subgrid reservoir of APE (turbulent potential
energy, TPE) have been proposed \citep{zilitinkevich_hierarchy_2013}.
In such models $\overline{b'w'}$ is a conversion term between TKE
and TPE. However their energy budget is relatively ad hoc and comes
with restrictions such as a constant Brunt-Vaisala frequency $N^{2}$.
It would be interesting to extend the present framework to include
an explicit representation of TPE and its conversion to/from TKE and
BPE, as this might provide additional constraints on closures.

LES models are able to represent convection explicitly, but coarser-grain
models have to represent it via closures. The locality assumption
central to the present work then breaks, at least for those turbulent
fluxes carried by convective plumes. Non-local approaches, especially so-called
mass-flux closures, have been proposed to express convective fluxes.
\citep{perrot_energetically_2025} have proposed such a closure with
a closed energy budget for simple thermodynamics. It would be interesting
to generalize \cite{perrot_energetically_2025} to arbitrary thermodynamics,
to identify its entropy production and to examine its consistency
with the second law. 

\section*{Acknowledgements}

This work is part of the AWACA project that has received funding from
the European Research Council (ERC) under the European Union's Horizon
2020 research and innovation program (grant agreement no. 951596).\\
This study has received funding from Agence Nationale de la Recherche
- France 2030 as part of the PEPR TRACCS programme under grant number
ANR-22-EXTR-0006.

The author would like to thank John Thuburn for his comments on an early version of this manuscript.

\bibliographystyle{jfm}
\bibliography{ConsistentLES}


\end{document}